\documentclass[reprint,
superscriptaddress,
amsmath,amssymb,
aps,
pra,
]{revtex4-2}
\setcitestyle{}
\usepackage{graphicx}
\usepackage{dcolumn}
\usepackage{bm}
\usepackage{dsfont}
\usepackage{physics}
\usepackage{bbold}
\usepackage{color}
\usepackage{xcolor}
\usepackage{soul}
\setstcolor{red}
\usepackage{ifthen}
\newboolean{showannotations}   
\setboolean{showannotations}{false} 
\newcommand*{\added}[1]{%
  \ifthenelse{\boolean{showannotations}}{{\color{blue}#1}}{#1}%
}

\newcommand*{\removed}[1]{%
  \ifthenelse{\boolean{showannotations}}{\st{#1}}{}%
}

\newcommand*{\change}[2]{%
  \ifthenelse{\boolean{showannotations}}{{\color{red}\st{#1}}{\color{blue}#2}}{#2}%
}

\begin{document}

\title{ Multiconfigurational Mixed Quantum-Classical Approach \\for Correlated Many-Body Dynamics}
\author{Pritha Ghosh}%
\affiliation{Department of Chemistry, Texas A\&M University, College Station, Texas 77843, USA}
\author{Rajanya Sarkar}%
\affiliation{Department of Chemistry, Texas A\&M University, College Station, Texas 77843, USA}
\author{Arkajit Mandal}%
\email{mandal@tamu.edu}
\affiliation{Department of Chemistry, Texas A\&M University, College Station, Texas 77843, USA}

\begin{abstract}
In this work, we introduce a multiconfigurational mixed quantum-classical many-body  approach for simulating the finite-temperature correlated multi-exciton dynamics in the presence of phonon-induced static and dynamic disorder. In this mixed quantum-classical approach, the excitonic subsystem is described using a multiconfigurational wavefunction that extends beyond the mean-field limit, while the phonons are evolved quasi-classically. Using this approach, we simulate a multi-excitonic dissipative system and show how the interplay between phonon-induced dynamic disorder and exciton-exciton many-body interactions determines excitation-dependent excitonic transport and spatial correlations. Our results show that while the mean-field approach produces semi-quantitatively accurate diffusive dynamics, it does not capture the spatial correlations as expected.  We find that a mean-field path approximation, where we generate pre-computed trajectories using our mean-field mixed quantum-classical approach and then perform multiconfigurational dynamics, can  reproduce the spatial correlations to a good accuracy, positioning this approach as an efficient method for capturing spatial correlations in complex systems. 
\end{abstract}
\maketitle
\section{Introduction}
Exciton dynamics is central to understanding energy transport in a diverse range of semiconductors and organic materials ~\cite{Dimitriev2022, Engel2007, Akselrod2014, Deng2020, Jin2018, Najafov2010, zhu2023many} as well as in the emerging field of strongly coupled light-matter hybrid systems~\cite{Balasubrahmaniyam2023, Khazanov2023, Krupp2025}.
A growing number of technological applications, such as organic photovoltaics~\cite{Guo2024, Price2022}, light-emitting diodes,~\cite{Baek2025, Giebink2008} and nonlinear optical devices including optical switches~\cite{Grosso2009}, modulators,~\cite{lee2024ultra} and frequency converters~\cite{PhysRevLett.132.246902}, harness the rich dynamical interplay of electronic, photonic, and phonon degrees of freedom. However, simulating excitonic dynamics, especially many-body phenomena such as biexciton formation~\cite{Sohoni2024}, exciton-exciton interactions and annihilation~\cite{Giebink2008, Deng2020, Kumar2023}, excitonic condensation~\cite{Butov2002, High2012, Wang2019}, Auger recombination~\cite{PhysRevLett.96.057407} and singlet fission~\cite{Berkelbach2013b, Tempelaar2018}, remains challenging because of the intrinsic many-body nature of the problem and the resulting non-adiabatic dynamics due to the coupling to phonon degrees of freedom. Such many-body effects are also of growing importance in strongly coupled light-matter systems~\cite{Balasubrahmaniyam2023, Khazanov2023, Krupp2025, amini2026many, ghosh2025mean}, where excitons hybridize with photons to form exciton-polaritons that can exhibit exotic quantum phenomena even at room temperature, making this an ideal platform for developing robust quantum technologies.

An exact quantum dynamical simulation for propagating the phonon-coupled correlated many-body exciton dynamics is computationally intractable. Meanwhile, mixed quantum-classical (MQC) formalisms~\cite{ LiJCTC2019, CrespoCR2018,TullyJCP1990, SubotnikARPC2016, LiJPCL2022, MandalCR2023}, where slow degrees of freedom are evolved using Hamilton's equation of motion and fast degrees of freedom are propagated quantum mechanically, provide a reasonable balance between accuracy and computational efficiency. However, an accurate incorporation of (multi-excitation) many-body effects within the MQC formalism is still a challenging task. Often, prior theoretical works confine the excitonic dynamics to the single excitation subspace~\cite{liu2015ligands, wang2011mixed, sneyd2022new, stippell2025computational, einsele2024nonadiabatic, troisi2006charge, Prezhdo2021ACSAcc, akimov2014nonadiabatic}, which, while keeping the simulation tractable, cannot capture the many-body effects as expected. At the same time, mean-field incorporation of many-body effects~\cite{morita2022observation, AlliluevJLTP2024, CarusottoRMP2013, NespoloPRB2019} also misses quantum correlations and thus is particularly unsuitable for studying applications in quantum technologies~\cite{zhu2023many}. This includes our recent work~\cite{ghosh2025mean}, which propagates the  many-body quantum dynamics of excitons (and exciton-polaritons) using a mixed quantum-classical approach where the many-body effects are included in a mean-field fashion. Consequently, our mean-field mixed quantum-classical approach cannot capture spatial quantum correlations in phonon-coupled excitons or exciton-polaritons, limiting its scope for {\it in silico} investigation of quantum materials for quantum information processing.

 In the present work, we address this challenge by developing a multiconfigurational mixed quantum-classical approach where the quantum subsystem is written as a linear combination of permanents, which allows for including many-body correlations in a systematic manner, while the phonon degrees of freedom propagate quasi-classically with non-adiabatic forces that are consistent with the correlated description of the quantum subsystem. This approach reduces to the multi-trajectory Ehrenfest (MTE)~\cite{TullyFD1998, KoshkakiNatCommun2026, BlackhamNanoLett2025,ghosh2025mean} formalism (on a single excited subspace) when we consider a single bosonic excitation ($N_\mathrm{ex}=1$). On the other hand, it reduces to the multiconfigurational time-dependent Hartree for bosons (MCTDHB)~\cite{AlonPRA2008,Streltsov2007,Meyer2009MCTDHReview} approach when ignoring the coupling to the phonon degrees of freedom. At the same time, we note that the present method shares no similarities to the `Multiconfigurational Ehrenfest' approach, which focuses on capturing quantum nuclear dynamics and not multi-excitonic effects.~\cite{shalashilin2011multiconfigurational}
 
 The multiconfigurational ansatz used here is a linear combination of all the participating many-body bosonic configurations (called permanents) for a certain total number of excitations ($N_\mathrm{ex}$), which is made up of a number of bosonic modes ($M$)~\cite{AlonPRA2008,Streltsov2007,Meyer2009MCTDHReview}. The variational manifold in this type of quantum propagation thus jointly consists of time-dependent bosonic modes, each of which may be occupied by one or more excitations, and the associated time-dependent coefficients of all possible many-body bosonic permanents. 
 The time-dependent modes and the coefficients of the permanents evolve in a coupled manner to generate the many-body quantum dynamics. Allowing progressively more modes increases the accuracy of the approach, which converges to the exact quantum propagation at a relatively few number of such modes. Our method provides a path to understanding the interplay between exciton-exciton interaction and phonon-induced dynamic disorder in spatially correlated systems. Although we observe only a modest difference in diffusive propagation between the single-mode and multimode approaches, the difference in the spatial pair density signature is significant. Further, we implement three specific approximations to the multiconfigurational phonon-coupled exciton dynamics and find that the mean-field path approximation (MFPA)  captures the early-time spatial correlation to good accuracy.  While out of the scope of the present work, we expect this new many-body quantum dynamical approach to be useful for understanding the correlated many-body dynamics of exciton-polaritons, which will be the subject of our future work.

 \section{Theory}
 {\bf Many-body exciton-phonon Hamiltonian.} In this work, we consider a Bose-Hubbard-Holstein Hamiltonian~\cite{ghosh2025mean}. As a representative model we consider an one-dimensional chain of an excitonic material that consists of $N$ sites with each site coupled to a phonon degree of freedom.
  The many-body Hamiltonian, which describes excitons coupled to phonons, is written as
 \begin{equation}\label{Hamiltonian-LM}
 \hat{H} = \hat{H}_\mathrm{e} + \hat{H}_\mathrm{p} + \hat{H}_\mathrm{e-p}. 
 \end{equation}
 Here, $\hat{H}_\mathrm{e}$ is the excitonic Bose-Hubbard Hamiltonian which includes on-site repulsive interactions and is written as
\begin{equation}
\hat{H}_\mathrm{e} = \sum^N_{i, j} \epsilon_{i,j} \hat{X}_{i}^\dagger \hat{X}_{j} +  \frac{U }{2}\sum^N_{i} \hat{X}_i^\dagger \hat{X}_i^\dagger \hat{X}_i \hat{X}_i,
\end{equation}
 where $\hat{X}_{i}^{\dagger}$ creates an exciton at a site $i$. Here, $\epsilon_{ij} = \epsilon_{0}\delta_{ij} - \tau (\delta_{i, j+1} + \delta_{i, j-1})$ refers to the one-body parameters, with $\epsilon_{0}$ as the on-site energy and $\tau$ as the hopping parameter. Finally, $U$ is the strength of the on-site repulsive interactions in our model system.

The phonon part of the Hamiltonian, i.e., $\hat {H}_\mathrm{p}$, is described by a set of harmonic oscillators and is written as
\begin{equation}
\hat{H}_\mathrm{p} = \sum^N_{i} \frac{\hat P_i^2}{2} + \frac{1}{2}\omega_0^2{\hat R_i^2}, 
\end{equation}
where $\{\hat P_i\}$ and $\{\hat R_i\}$ are the corresponding momentum and position operators respectively, for the phonon degrees of freedom that couple to the excitons, while the phonon frequency is $\omega_0$. The exciton-phonon coupling is written as
\begin{equation}\label{ep-coupling}
\hat{H}_\mathrm{e-p}= \gamma \sum^N_{i}  \hat{X}_{i}^\dagger\hat{X}_i  \hat{R}_i ,
\end{equation}
where $\gamma$ is the strength of the exciton–phonon coupling.

{\bf Multiconfigurational Many-body Ehrenfest Approach.} In this approach, the nuclear degrees of freedom are treated quasi-classically ($\{\hat{R}_i, \hat{P}_i\} \rightarrow \{{R}_i, {P}_i\} $). Meanwhile, the excitonic degrees of freedom evolve in a quantum mechanical fashion. We adopt the following many-body ansatz:
\begin{equation}\label{ansatz}
|\Psi(t)\rangle=\sum_{\vec{n}}C_{\vec{n}}(t)|\vec{n};t\rangle~.
\end{equation}
 Here, \{$C_{\vec{n}}(t)$\}s are the corresponding time-dependent coefficients of the time-dependent many-body permanents \{$|\vec{n};t\rangle$\}, where
\begin{align}\label{permanent expansion}    
|\vec{n};t\rangle=&\frac{(\hat {B}_{M-1}^{\dagger}(t))^{n_{M-1}}\hdots(\hat {B}_1^{\dagger}(t))^{n_1}(\hat {B}_0^{\dagger}(t))^{n_0}}{\sqrt{n_0!n_1!\hdots n_{M-1}!}}|\bar{0}\rangle\nonumber\\
=&{\prod_{k=0}^{M-1}}\frac{(\hat {B}_k^{\dagger}(t))^{n_k}
}{\sqrt{n_k!}}|\bar{0}\rangle.
\end{align}
Here, $|\bar{0}\rangle$ denotes the vacuum state. The explicit time dependence of the permanents originates from the time-dependent nature of the modes, i.e., \{$\hat {B}_k^{\dagger}(t)$\} ($k=0,1,2,3,...,M-1$).  The size of the full space of all possible many-body permanents is given as $\binom{N_\mathrm{ex}+M-1}{N_\mathrm{ex}}$, where $N_\mathrm{ex}$ refers to the total number of bosonic excitations (excitons) in the system, and it remains conserved throughout the dynamics. Further, $M$ is the total number of  modes chosen to reproduce the many-body dynamics. The operator $\hat {B}_k^{\dagger}(t)$ ($\hat {B}_k(t)$) creates (annihilates) a bosonic excitation in mode $k$, corresponding to the time-dependent single particle function (SPF)  $|\phi_k(t)\rangle$. Specifically, $\hat {B}_k^{\dagger}(t)$ is related to $|\phi_k(t)\rangle \equiv [\phi_{k0}(t), \phi_{k1}(t), ... \phi_{kj}(t), ...]$ as 
\begin{align}
\hat {B}^{\dagger}_k(t)=\sum^N_j\phi_{kj}(t)\hat{X}_j^{\dagger},
\end{align}
where $\phi_{kj}(t)=\langle j| \phi_{k}(t)\rangle$; $j$ denotes the discrete site index ($j=0,1,2...,N-1)$ and $k$ denotes the usual mode index. \{$\hat {B}_k(t)$\} follows the typical bosonic commutation relations as \{$\hat {X}_j(t)$\}. We note that the permanent $|\vec{n};t\rangle$ has a simple form in occupation number representation, i.e., $|\vec{n};t\rangle= |n_0,n_1,...,n_{M-1};t\rangle$, where $n_0+n_1+...+n_{M-1}=N_\mathrm{ex}$ is the restriction coming from the confinement to the $N_\mathrm{ex}^{th}$ excitation subspace.
Following the standard Dirac–Frenkel formulation of the time-dependent variational principle, which uses the variational condition $\frac{\partial}{\partial \xi^*}\left(\langle\Psi(t)|\hat{H}-i\frac{\partial}{\partial t}|\Psi(t)\rangle\right)=0$, where $\xi^* \in \{ \phi^*_{ki}(t),C^*_{\vec{n}}(t) \}$, we obtain the coupled equations of motion (EOMs) for the discrete site-basis representations of the SPFs (\{${\phi}_{ki}(t)$\})  and the expansion coefficients (\{$C_{\vec{n}}(t)$\}) as
\begin{align}\label{EOM1}
i\dot{\phi}_{ki}(t) = ~&\hat{\mathrm{Q}}(t) \Bigg[\sum^{N}_{j}h_{ij}\phi_{kj}(t)\\
+~ U\sum^M_{q,b,c,d}&\left(\rho^{(1)}(t)\right)^{-1}_{kq}\rho^{(2)}_{qcdb}(t)\phi^*_{ci}(t)\phi_{di}(t)\phi_{bi}(t)\Bigg],\nonumber
\end{align}
and 
\begin{align}\label{EOM2}
i\dot{C}_{\vec{n}}(t)=\sum_{\vec{n}'}C_{\vec{n}'}(t)\hat{\mathcal{H}}_{\vec{n}\vec{n}'}.
\end{align}
In Eqn.~\ref{EOM1}, $h_{ij} = \langle \bar{0}|\hat{X}_{i}\hat{H}_{(1)}\hat{X}^\dagger_{j}| \bar{0}\rangle$ is the matrix element of the one-body Hamiltonian written in the site basis, where $\hat{H}_{(1)}= \hat{H}-\hat{H}_p-\frac{U }{2}\sum_{i} \hat{X}_i^\dagger \hat{X}_i^\dagger \hat{X}_i \hat{X}_i$, and $\rho^{(1)}$ is the one-body reduced density matrix, whose element in SPF (or mode) basis is expressed as 

\begin{align}
\rho^{(1)}_{kq}(t)= \sum_{\vec{n},\vec{n}'}C^*_{\vec{n}}(t)C_{\vec{n}'}(t)\langle\vec{n};t|\hat {B}_k^{\dagger}(t)\hat {B}_q(t)|\vec{n}';t\rangle,
\end{align}
and $\rho^{(2)}$ is the corresponding two-body reduced density matrix, whose matrix element is written as 
\begin{align}
\rho^{(2)}_{q,b,c,d}(t) &= \sum_{\vec{n},\vec{n}'}C^*_{\vec{n}}(t)C_{\vec{n}'}(t) \nonumber\\ 
&\times \langle\vec{n};t|\hat {B}_q^{\dagger}(t)\hat {B}_b^{\dagger}(t)\hat {B}_c(t)\hat {B}_d(t)|\vec{n}';t\rangle.
\end{align}

\begin{figure*}[!t]
\centering
\includegraphics[width=1.0\linewidth]{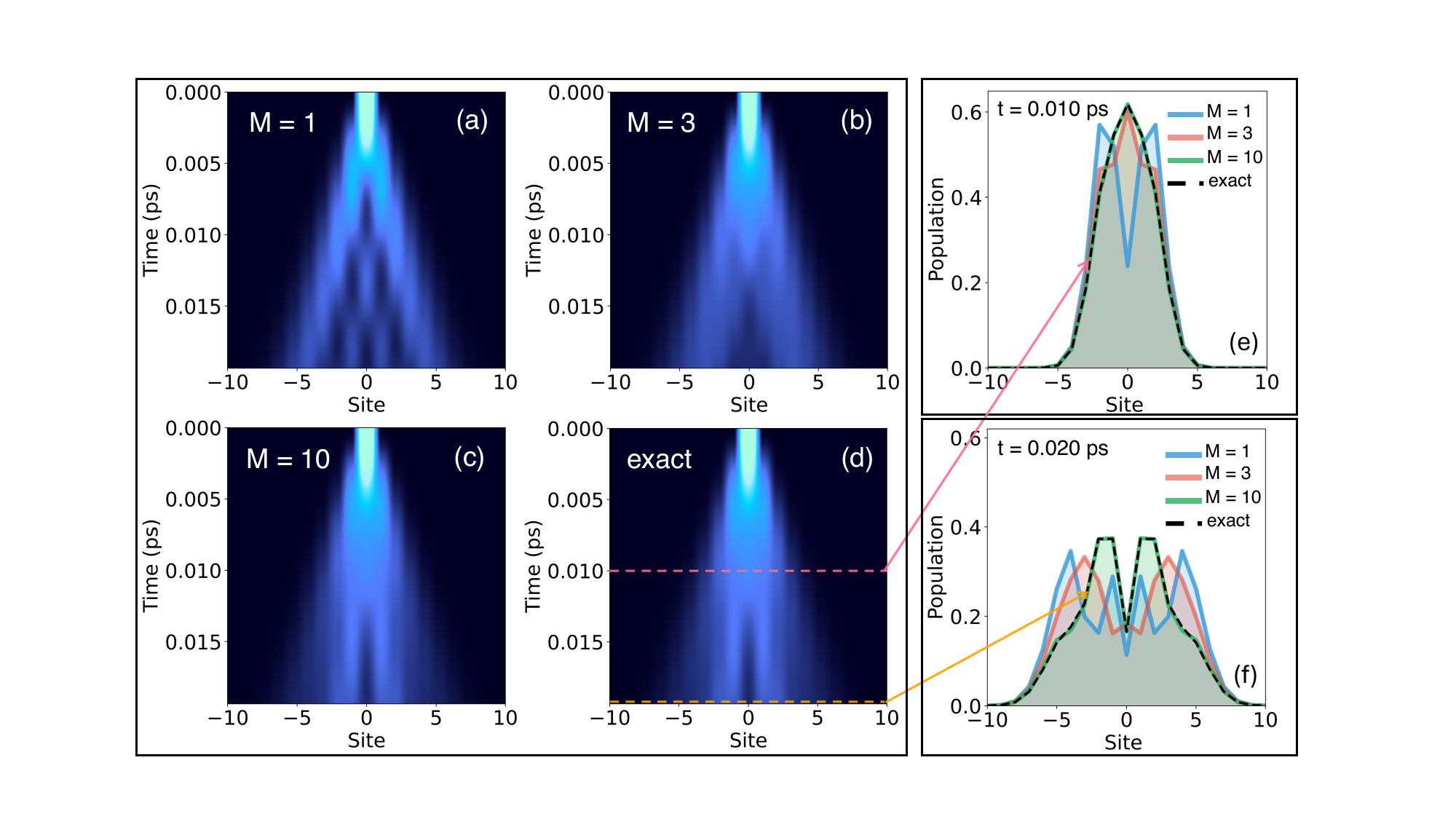}
\caption{\footnotesize 
Multiconfigurational many-body dynamics of a centrally initialized  bosonic density of 3 excitations over a lattice of 21 sites, compared with the exact (Schr\"odinger) propagation of the same system. (a)-(c) The single-mode ($M=1$), three-mode ($M=3$) and ten-mode ($M=10$) dynamics, respectively, of the bosonic density over time (in picoseconds). (d) The exact dynamics of the same 3-excitation system. (e)-(f) The spatial population profiles of the bosonic density at time $0.010$ ps and $0.020$ ps, respectively, for the representative cases ($M =1,3,10$) (shown in (a)-(c)) and the exact (Schr\"odinger) propagation (shown in (d)). The nearest-neighbor hopping parameter ($\tau$) is set to 0.004 atomic units (a.u.), while the interaction strength ($U$) is $0.006$ a.u. }
\label{mbfig1}
\end{figure*}

In order to prevent numerical instabilities from nearly unoccupied SPFs in our algorithm, we evaluate the inverse of the one-body reduced density matrix using a regularization scheme in which only the SPFs with occupations exceeding a given threshold contribute to the inverse, i.e.,

\begin{align}
\left(\rho^{(1)}(t)\right)^{-1} \simeq  \hat{V}\mathcal{P}\Bigg[ \sum^{M-1}_{k=0} \frac{1}{n_k(t)}|\tilde{\phi}_k(t)\rangle \langle \tilde{\phi}_k(t)|\Bigg]\mathcal{P}\hat{V}^{\dagger} ,
\end{align}
where $\hat{V}$ is a unitary matrix that diagonalizes $\rho^{(1)}(t)$, $\{n_k(t)\}$ are the eigenvalues and $\mathcal{P} = \sum_{n_k>\varepsilon}|\phi_k(t)\rangle \langle \phi_k(t)|$ projection operator with $\varepsilon$ a small numerical threshold.

In Eqn.~\ref{EOM2}, $\hat{\mathcal{H}}_{\vec{n}\vec{n}'}=\langle \vec{n};t|\hat{\mathcal{H}}|\vec{n}';t\rangle$ refers to the matrix element of the many-body Hamiltonian $\hat{\mathcal{H}}= \hat{H}-\hat{H}_\mathrm{p}$.\\
To tackle the originally existing non-uniqueness in the solutions of the EOMs (as is done in MCTDHB~\cite{sakmann2011many}), we employ a time-dependent one-body projection operator $\hat{\mathrm{Q}}(t)$ to perform a gauge fixing on the EOM of the SPFs, where
\begin{align}
\hat{\mathrm{Q}}(t)=1-\sum^{M-1}_{k'=0}|\phi_{k'}(t)\rangle\langle\phi_{k'}(t)| 
\end{align}
maintains the condition that the time-derivative of $|\phi_{k}(t)\rangle$ for any $k$ should not have contributions from the current \{$|\phi_{k}(t)\rangle$\}s themselves, i.e., $\langle \phi_k(t)|\dot{\phi}_q(t)\rangle =0, ~\forall ~k,q \in \{0,1,2,...,M-1\}$, leading to the current Eqn.~\ref{EOM1} (see details in the Supplemental Material). This condition mathematically preserves the orthonormality of the SPFs at all times. However, the numerical implementation still causes some drift in orthonormality, which is externally handled by an infrequent explicit re-orthonormalization and corresponding gauge-transformation of the set of coefficients \{$C_{\vec{n}}(t)$\} to match the orthonormalized SPFs. Specifically, an orthonormalization process on the SPFs is expressed as a linear transformation 
\begin{align}
|\phi_i'(t)\rangle = \sum_{j=0}^{M-1}  U_{ij}|\phi_j(t)\rangle,
\end{align}
where \{$|\phi_i'(t)\rangle$\} is the transformed set of SPFs. This causes the many-body permanents to transform as  
\begin{equation}
|\vec n(\phi');t\rangle
=
\sum_{\vec m}
\sqrt{\prod_{j=0}^{M-1} m_j!}\,
A_{\vec m,\vec n}
\, |\vec m(\phi);t\rangle,
\end{equation}
where,
\begin{equation}
A_{\vec m,\vec n}
=
\sum_{\{k_{ij} \to \vec m\}}
\prod_{i=0}^{M-1}
\frac{n_i!}{\prod_{j=0}^{M-1} k_{ij}!}
\prod_{j=0}^{M-1} U_{ij}^{k_{ij}}.
\end{equation}
Here, $\vec m,\vec n$ denote the indices of the permanents, and the occupations are partitioned as 
\begin{equation}
m_j = \sum_{i=0}^{M-1} k_{ij}.
\end{equation}
Hence, the corresponding set of coefficients gets transformed according to the expression
\begin{equation}
C'_{\vec m}(t)
=
\sum_{\vec n}
A'_{\vec m,\vec n}\, C_{\vec n}(t),
\end{equation}
where,
\begin{equation}
A'_{\vec m,\vec n}
=
\sqrt{\prod_{j=0}^{M-1} m_j!}
~A_{\vec m,\vec n}.
\end{equation}

 Finally,  we obtain the nuclear equation of motion using the multiconfigurational ansatz in Eq.~\ref{ansatz}, written as 
{
\begin{align} \label{nuclear-EOM}
\dot{R}_i(t) &= P_i(t)\nonumber \\ 
\dot{P}_i(t) &= -\omega_0^2R_i(t) - \Big\langle \Psi(t)\Big|\frac{d \hat{H}_\mathrm{e-p}}{dR_i}\Big|\Psi(t)\Big\rangle \nonumber \\
&= -\omega_0^2R_i(t) - \gamma\sum_{kq}\phi^*_{ki}(t)\phi_{qi}(t)\rho^{(1)}_{kq}(t).
\end{align}
}

Here, $\sum_{kq}\phi^*_{ki}(t)\phi_{qi}(t)\rho^{(1)}_{kq}(t) = \rho^{(1)}_{ii}(t)$ is the spatial probability density at site index $i$, which indicates that excitations in all accessible modes contribute to the estimation of the mean force that drives each nucleus. We note that, in the absence of coupling to phonons, the whole treatment essentially becomes purely MCTDHB evolution~\cite{AlonPRA2008,Streltsov2007,Meyer2009MCTDHReview}.

\section{Results and Discussion}

\begin{figure*}[!t]
\centering
\includegraphics[width=1.0\linewidth]{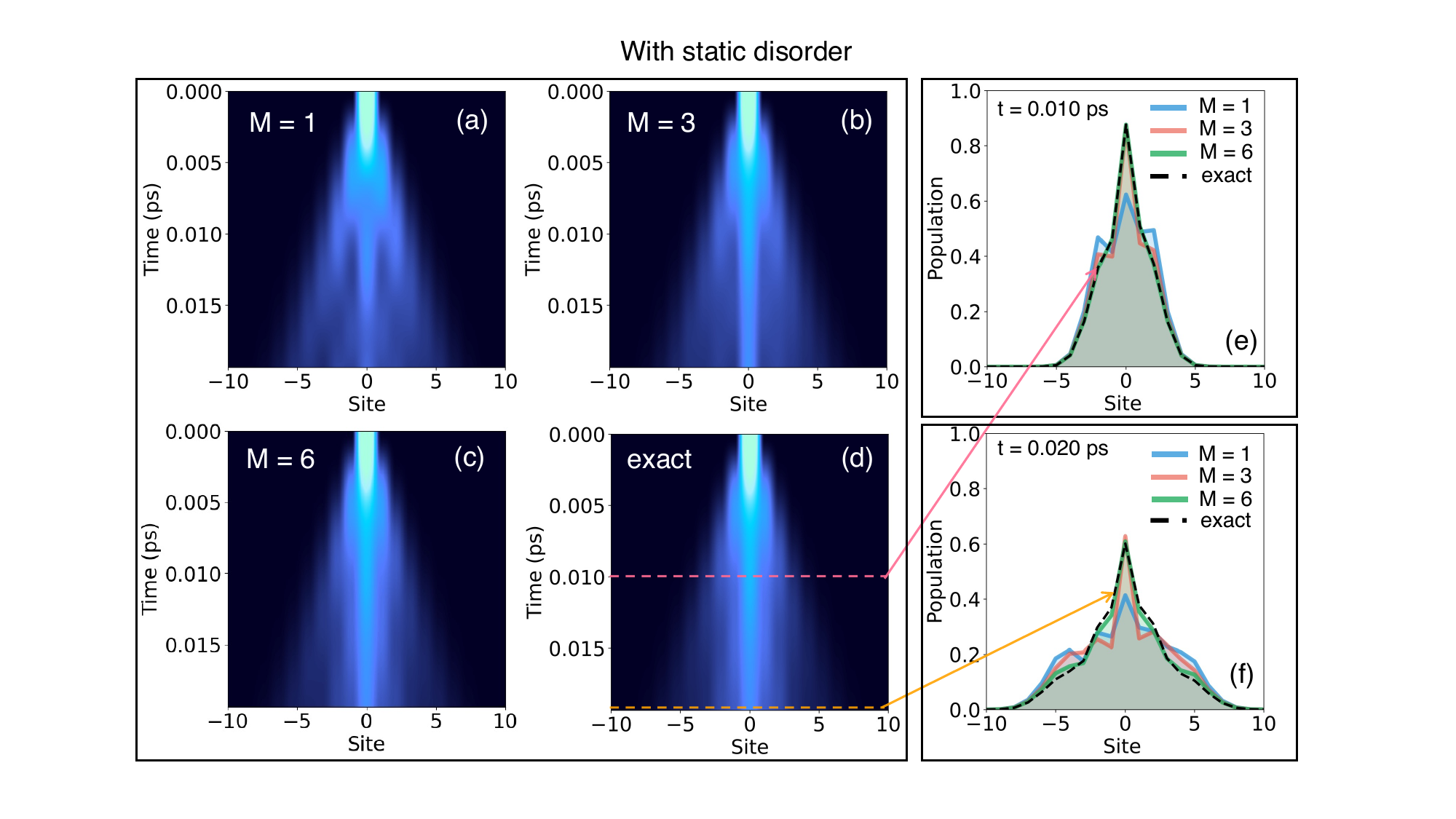}
\caption{\footnotesize 
Multiconfigurational many-body dynamics of a centrally excited 3-excitation wave packet in the presence of 2\% static disorder, where the on-site energies are sampled from a normal distribution. (a)-(c) The single-mode ($M=1$), three-mode ($M=3$) and six-mode ($M=6$) dynamical evolution, respectively, of the bosonic density over time (in ps). (d) The exact dynamics of the same 3 -excitation system (for the same realizations of static disorder as in (a)-(c)). (e)-(f) Disorder-modified spatial population profiles of the bosonic density at time 0.010 ps and 0.020 ps, respectively, for the representative cases ($M =1,3,6$) (shown in (a)-(c)) and the case of the exact quantum propagation (shown in (d)). The nearest-neighbor hopping parameter ($\tau$) is set to $0.004$ a.u., while the interaction strength is $U=0.006$ a.u. }
\label{mbfig2}
\end{figure*}
{\bf Convergence of the multiconfigurational approach towards exact dynamics.}
Fig.~\ref{mbfig1} shows the pure multiconfigurational many-body quantum dynamics of a bosonic density, which is initialized as a delta function at the center of the 21-site model lattice (in the absence of phonons), to benchmark the method against the exact quantum (Schr\"odinger, i.e., $i|\dot{\Psi}(t)\rangle=\hat{H}|\Psi(t)\rangle$) propagation for the same system. Fig.~\ref{mbfig1}(a)-(c) illustrate the symmetric evolution of the spatial bosonic density ($\rho^{(1)}_{ii}$), for different numbers of modes: single-mode ($M = 1$), three-mode ($M = 3$), and ten-mode ($M = 10$) cases, respectively. Fig.~\ref{mbfig1}(d) presents the exact (Schr\"odinger) dynamics of the same 3-excitation system. The many-body ansatz for the exact dynamical evolution is initialized as
\begin{align}
|\Psi(t=0)\rangle =  \bigotimes_{i=1}^{N}|n_i;(t=0)\rangle,
\end{align}
where $i$ is the site index; $n_i(t=0)= N_\mathrm{ex} \delta_{i,(N+1)/2}$ for an odd number of sites and $n_i(t=0)= N_\mathrm{ex} \delta_{i,N/2}$ for an even number of sites. Here we use $N_\mathrm{ex} = 3$.

Meanwhile, the equivalent initialization used in the multiconfigurational dynamics is represented as
\begin{align}
|\Psi(t=0)\rangle = \sum_{\vec{n}}C_{\vec{n}}(t=0)|\vec{n};t=0\rangle,
\end{align}
where $C_{\vec{n}}(t=0) =1$, only for $\vec{n}=(N_\mathrm{ex},0,0...,0)$ and $0$
otherwise. This indicates that the entire population of the excitations is present in the first SPF ($|\phi_0\rangle$), whose distribution in space is a delta function at the central site. The normalized spatial profiles of the rest of the accessible SPFs ($\{|\phi_1\rangle, |\phi_2\rangle,...,|\phi_{M-1}\rangle\}$) are defined such that they are orthogonal to each other and to $|\phi_0\rangle$. Although these other SPFs are unoccupied initially, the multiconfigurational evolution allows the overall bosonic population to be redistributed among all the SPFs accordingly (unless $M=1$).

Fig.~\ref{mbfig1}(e)-(f) show the spatial population distribution at an intermediate time ($0.010$ ps) and at  $0.020$ ps for all the representative cases ($M=1,3,10$), along with the exact quantum propagation. The nearest-neighbor hopping parameter ($\tau$) is $0.004$ atomic units (a.u.), and the interaction strength ($U$) is $0.006$ a.u. The results illustrate that as the number of modes ($M$) included in the multiconfigurational approach increases, the dynamics progressively converge toward the exact quantum dynamics, as expected.\\

{\bf Convergence of the multiconfigurational dynamics in the presence of static disorder.} Fig.~\ref{mbfig2} presents the multiconfigurational dynamics of a bosonic system consisting of 3 excitations, where the on-site energies are sampled from a normal distribution with a standard deviation of $2\%$ of the mean on-site energy ($\sigma_{\epsilon}=0.02\epsilon_{0}$), and the results are averaged over 40 different disorder realizations. This 3-excitation system is initialized centrally, with its spatial population distribution given by a delta function (same as in Fig.~\ref{mbfig1}).

\begin{figure*}[!t]
\centering
\includegraphics[width=1.0\linewidth]{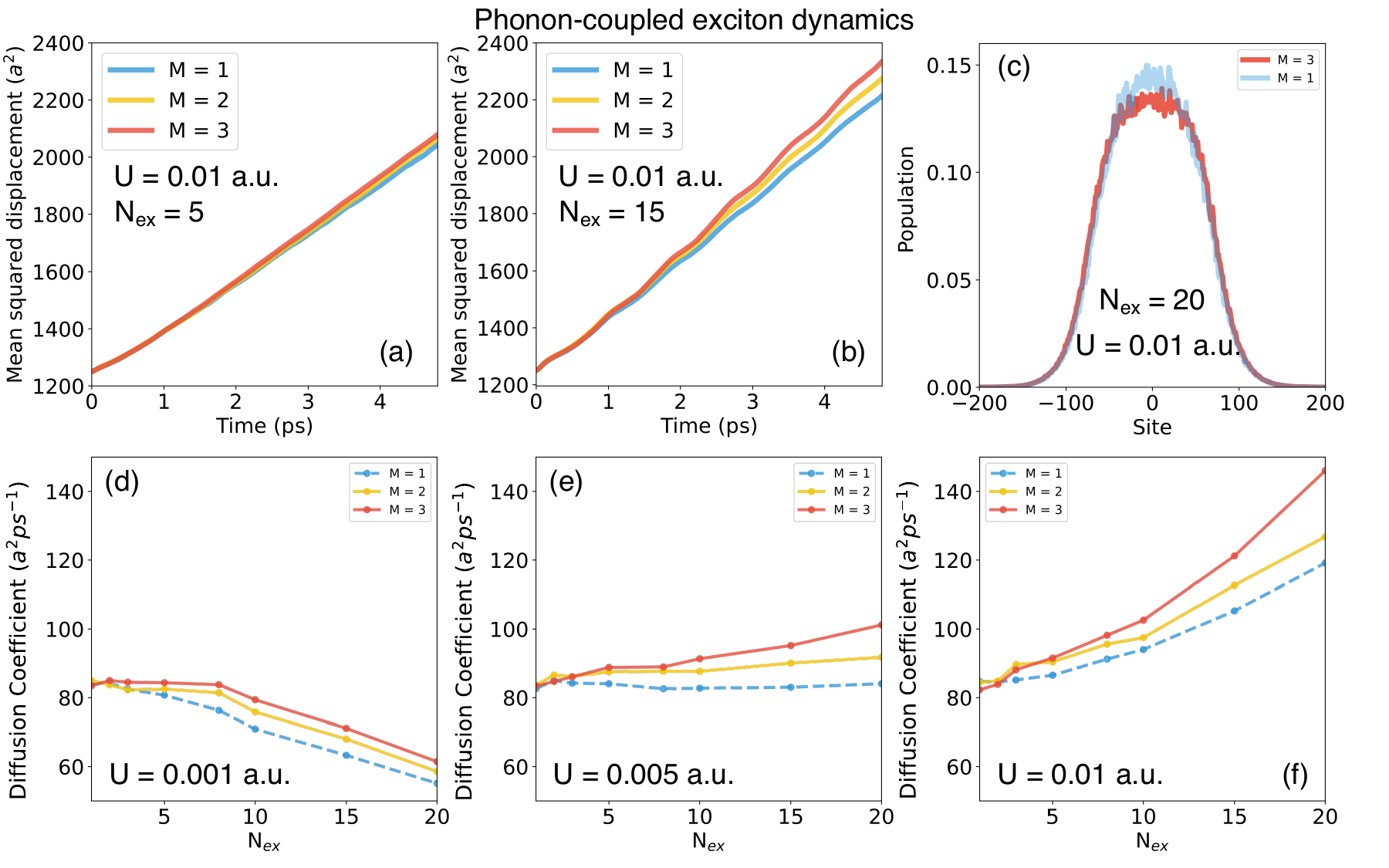}
\caption{\footnotesize 
Phonon-coupled exciton dynamics in the presence of dynamic disorder on a lattice of 400 sites at $150$ K. The excitonic density is initialized as a many-body Gaussian wave packet centered at the middle of the lattice with $\sigma=50$ units. (a)-(b) The mean squared displacement (MSD) of the excitonic density versus time (in ps) for single-mode ($M =1$) (solid blue line) as well as for multiconfigurational calculations: ($M = 2$) (solid yellow line) and ($M = 3$) (solid red line) at $U = 0.01$ a.u., for 5 and 15 excitations, respectively. (c) Final-time ($t_f=4.8$ ps) spatial excitonic density profiles for the $M=1$ (solid light blue) and $M=3$ (solid red) dynamics for a density initialized with 20 excitations. (d)-(f) Diffusion coefficient as a function of the number of excitations ($N_\mathrm{ex}$), for $M =1$ (dashed blue line), $M = 2$ (solid yellow line) and $M = 3$ (solid red line), at $U = 0.001$ a.u., $0.005$ a.u., and $0.01$ a.u., respectively. MSD is shown in units of $a^2$, where $a$ is the lattice constant. The nearest-neighbor hopping parameter ($\tau$) is set to $300~cm^{-1}$.}
\label{mbfig3}
\end{figure*}

Fig.~\ref{mbfig2}(a) shows the spreading out of the excitonic density over a net timespan of 0.02 ps when considering a single mode ($M=1$). Systematically increasing $M$ from $1$  to $3$ and then to $6$ (Fig.~\ref{mbfig2}(a)-(c)) shows that the dynamics gradually converge to that of the exact Schr\"odinger propagation of 3 excitations (Fig~\ref{mbfig2}(d)). There is an enhanced localization of excitonic density, in comparison to the case of no disorder (Fig.~\ref{mbfig1}). This occurs in response to the static disorder (which leads to Anderson localization) suppressing the spreading of the density by breaking the translational symmetry of the lattice. We find that a lower $M$ (here, $M=6$) is sufficient to obtain nearly convergent dynamics, unlike Fig.~\ref{mbfig1}, where a relatively higher $M$ is required to achieve convergence.  In addition,
Fig.~\ref{mbfig2}(e)-(f) compare the excitonic population profiles at two different times, respectively. At an intermediate time ($0.01$ ps), the $M=6$ dynamics is essentially converged, but at longer times, such as $0.02$ ps, we find a minor deviation from the excitonic population profile for the exact dynamics. Thus, the reliability of the results obtained from our multiconfigurational many-body method is not adversely affected when static disorder is incorporated, thereby establishing a reasonable foundation for extending this robust method to include dynamic disorder through the multiconfigurational mean field Ehrenfest approach. 

\begin{figure*}[!t]
\centering
\includegraphics[width=1.0\linewidth]{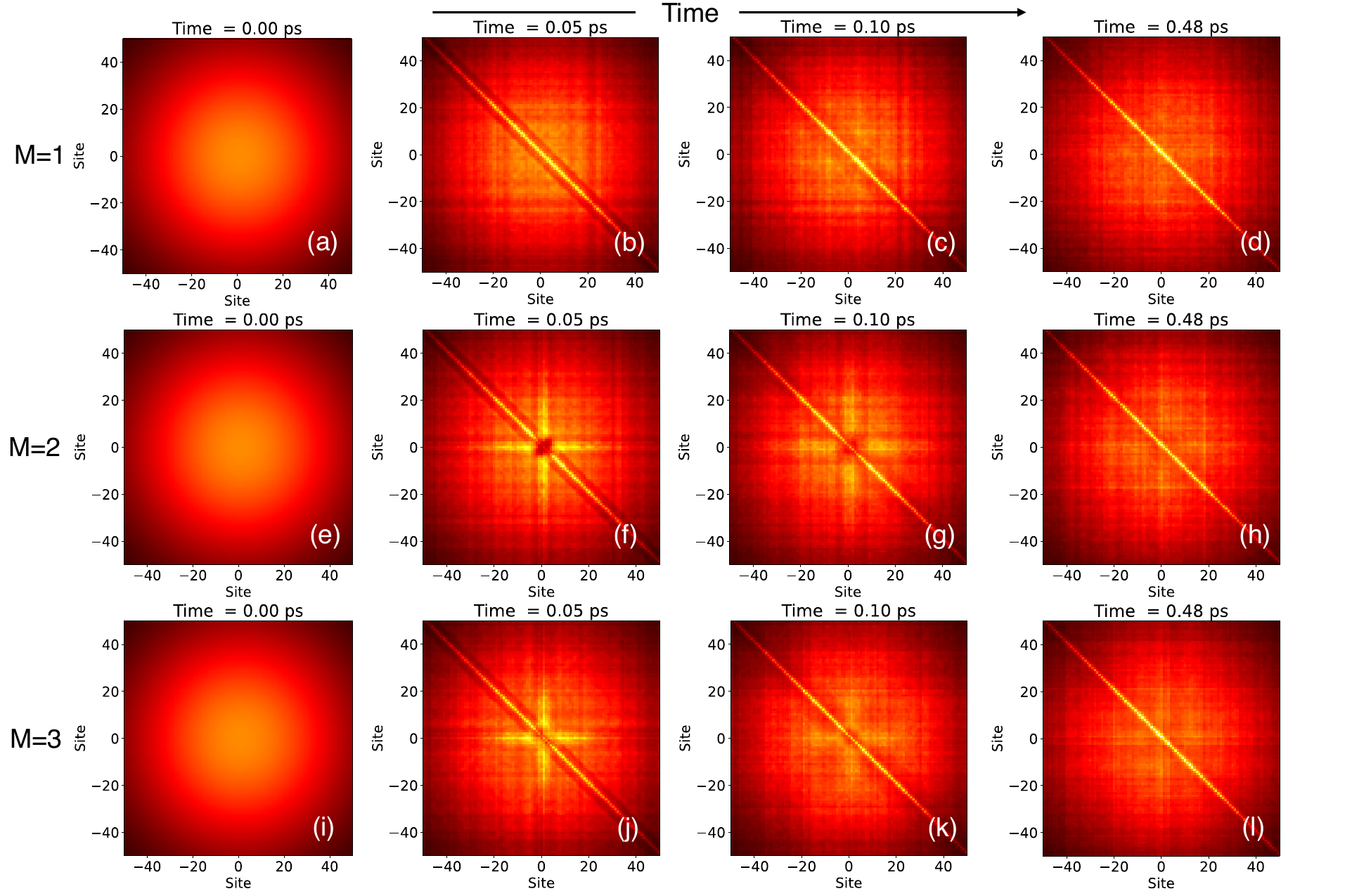}
\caption{\footnotesize 
Spatial pair-probability density, at different representative times (in ps) in the many-body dynamics of excitons at a temperature of $150$ K , for an initial 10-excitation wave packet on a lattice of 400 molecular sites. All the figures have been zoomed in to focus on the central zone where the dynamics hold relevance. (a)-(d) Snapshots of the single-mode $(M=1)$ evolution of the spatial two-body reduced density matrix ($\rho^{(2)}({i,j})$) at 0 ps, 0.05 ps, 0.10 ps and 0.48 ps respectively. (e)-(h) Snapshots of the two-mode $(M=2)$ dynamical evolution of $\rho^{(2)}(i,j)$ at the same representative times as in (a)-(d), capturing transient correlations. (i)-(l) Snapshots of the three-mode $(M=3)$ dynamical evolution of $\rho^{(2)}(i,j)$ at the same representative times as in (a)-(d), showing similar pair-density features that delocalize faster than (e)-(h). All the parameters used are the same as in Fig.~\ref{mbfig3}. }
\label{mbfig4}
\end{figure*}
{\bf Phonon-induced dynamical disorder in many-body exciton dynamics.} Fig.~\ref{mbfig3} demonstrates how the incorporation of dynamic disorder induced by phonon fluctuations affects the excitonic transport, namely the mean squared displacement (MSD)~\cite{wang2011mixed, sneyd2022new, GianniniNatCommun2022, ghosh2025mean} differently, depending on the number of modes ($M$) included in simulating the many-body quantum dynamics. The time-dependent MSD is defined as 
\begin{align}\label{MSD}
    \mathrm{MSD}(t) =\frac{1}{N_\mathrm{ex}} \sum^N_i \left(r_i - {r_\mathrm{com}(t)}\right)^2 \rho^{(1)}_{ii}(t).
\end{align}
Here, $r_\mathrm{com}(t)$ is the position of the center of mass of the many-body density at time $t$, evaluated as
\begin{align}\label{rcom}
   r_\mathrm{com}(t)= \frac{1}{N_\mathrm{ex}}\sum^N_i r_i\rho^{(1)}_{ii}(t). 
\end{align}
The phonon frequency corresponding to the nuclear vibration, i.e., $\omega_0$, is $5$ meV and the exciton-phonon coupling is set to $\gamma = 1.24\times 10^{-5}$ a.u., which is typical of crystalline organic semiconductors~\cite{wang2011mixed, troisi2006charge}. The temperature ($T$) is set to $150$ K. The  initial positions and momenta  of all 400 nuclei are sampled from a Wigner distribution with $\langle R \rangle =0, \langle P \rangle =0$, and
\begin{align}
\sigma_P
=
\frac{\sqrt{\omega_0}}
{\sqrt{2\,\tanh\left(\dfrac{\omega_0}{2k_BT}\right)}}
,\nonumber
~~\sigma_R
= \sigma_P/\omega_0.
\end{align}
We note that, for the range of temperature used in this work ($T>100$ K), the respective Wigner distributions of nuclear positions and momenta are nearly identical to the corresponding classical Maxwell-Boltzmann distributions. Thus, sampling from a Boltzmann distribution yields identical results. We consider the initial many-body excitonic density as a Gaussian centered at the middle site of the lattice (indexed by $i_{mid}=N/2$ if $N$ is even, or by $i_{mid}=(N+1)/2$ if $N$ is odd), with a spread of $\sigma=50$ lattice units, where the entire population occupies only $|\phi_{0}\rangle$, such that
\begin{align}
\phi_{0i}(t=0)= \frac{1}{\sqrt{\mathcal{N}}} \exp[{-\frac{(i-i_{mid})^2}{2\sigma^2}}].
\end{align}
Here $\mathcal{N}$ is the normalization constant. The rest of the SPFs are initialized such that they remain orthogonal. 

Fig.~\ref{mbfig3}(a)-(b) shows that the overall diffusive transport  is enhanced for 15 excitations ($N_\mathrm{ex}=15$) (Fig.~\ref{mbfig3}(b)) compared to 5 excitations ($N_\mathrm{ex}=5$) (Fig.~\ref{mbfig3}(a)), when the interaction strength $U$ is set to a moderately large value of $0.01$ a.u. More importantly, the difference in the transport dynamics of the excitonic density among the single-mode ($M=1$) and multi-mode ($M=3$) propagation becomes much more pronounced at a higher number of excitations. 

Fig.~\ref{mbfig3}(c) displays the spatial excitonic density profile for $N_\mathrm{ex}  = 20$ at $t = 4.8$ ps when using $M=1$ (mean-field limit) and $M=3$. We observe that the extent of the propagation is higher for the case of $M=3$, indicating that a mean-field ansatz underestimates the delocalization of the excitonic density, although the difference is marginal.

Fig.~\ref{mbfig3}(d)-(f) present the excitation number dependent diffusion coefficients for $M=1$ (dashed blue line), $M=2$ (solid yellow line), and $M=3$ (solid red line), with increasing many-body interaction strengths: $U=0.001$, $0.005$, and $0.01$ (in a.u.). The diffusion coefficient is calculated as 
\begin{align}
  D= \lim_{t\rightarrow t_\infty}\frac{1}{2} \frac{d}{dt}(\mathrm{MSD}(t)).  
\end{align}
In the case of low interaction strength ($U=0.001$ a.u.), the diffusion coefficients for all  different $M$s show an overall decreasing trend (Fig.~\ref{mbfig3}(d)), with increasing $N_\mathrm{ex}$. This indicates that having a higher number of excitations enhances the localization caused by phonon-induced dynamic disorder, resulting in reduced diffusion which is similar to what we found in our earlier work~\cite{ghosh2025mean}. However, when many-body interactions are more dominant (as shown by a higher interaction strength of $U=0.005$ a.u. in Fig.~\ref{mbfig3}(e)), it results in an opposite trend, where the diffusion coefficients increase with increasing the number of excitations. This is due to the increasing dominance of repulsive many-body interactions which hinders the localization caused by phonon-induced dynamical disorder. The diffusion coefficients are shown to increase further when using an even higher interaction strength (i.e., $U=0.01$ a.u.). 
\begin{figure*}[!t]
\centering
\includegraphics[width=1.0\linewidth]{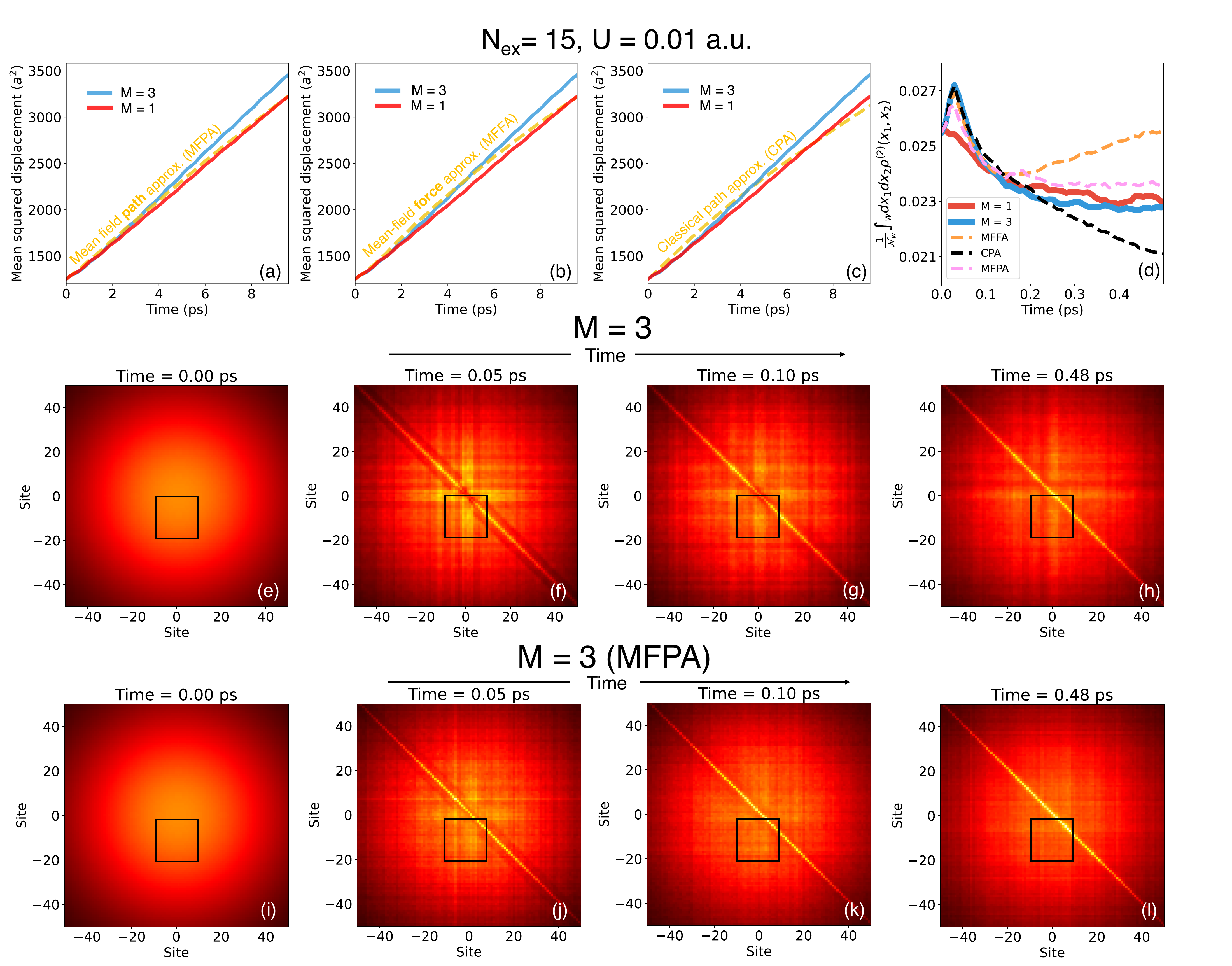}
\caption{\footnotesize 
Comparison of the mean squared displacement (MSD) for the multiconfigurational $M=3$ dynamics over time, where the MSD plot of the $M=1$ case is present as a reference, among three distinct types of approximations in the multiconfigurational ($M=3$) many-body Ehrenfest dynamics of a 15-exciton system with interaction strength $U=0.01$ a.u., followed by spatial pair density signatures over time in the approximate early dynamics that are compared to the reference $M=3$ dynamics. (a) MSD versus time plot for $M=3$ dynamics with precomputed positions and momenta (from $M=1$ dynamics) (mean field path approximation or MFPA). (b) MSD versus time plot for $M=3$ dynamics, but with the effective mean force on the nuclei generated from single-mode ($M=1$) dynamics, on the fly (mean field force approximation or MFFA). (c) MSD versus time plot for $M=3$ dynamics, where the nuclear force has no contribution from the quantum propagation (classical path approximation or CPA). (d) Average pair probability density $\rho^{(2)}$ calculated by considering a fixed square window ($W$) in the two-body coordinate space, having half-width $=5$ units and centered at $(0,-10)$ (indicated by a black square enclosure in the latter correlation maps) versus time (in ps), comparing the differences among the three approximations and the reference dynamics. $\mathcal{N}_W$ is the total number of points included within the window. (e)-(h) Snapshots of the reference three-mode $(M=3)$ dynamical evolution of $\rho^{(2)}(i,j)$ at different representative times: 0 ps, 0.05 ps, 0.10 ps and 0.48 ps, respectively. (i)-(l) Snapshots of the three-mode $(M=3)$ dynamical evolution  of $\rho^{(2)}(i,j)$ at the same representative times, under the mean field path approximation (MFPA). The rest of the parameters are identical to those in Fig.~\ref{mbfig3}.}
\label{mbfig5}
\end{figure*}

Fig.~\ref{mbfig3}(d)-(f)  show a modest difference between the single-mode and multi-mode dynamics, establishing that a single-mode ansatz (i.e., Gross-Pitaevskii ansatz) is somewhat insufficient to capture the correct dynamical propagation, while $M=2$ or $M=3$ dynamics systematically improves the excitonic propagation. We note that, while the nonadiabatic force expression does not directly depend on the exciton-exciton interaction strength $U$, the evolution of $\rho^{(1)}_{ii}(t)$ inherently involves the many-body contribution, which is explicitly present in the propagation of the SPFs, resulting in measurable differences in exciton transport.

Fig.~\ref{mbfig4} presents the time-dependent spatial pair densities, providing a qualitative visualization of the development of spatial correlations~\cite{CheneauNature2012}. Here, we set  $N_\mathrm{ex} = 10$. These results show that the emerging spatial pair density signatures decay due to the phonon-induced dynamical disorder and decoherence. The plots represent snapshots (at $t=0,~0.05,~0.10$ and $0.48$ ps, ordered from left to right) of the time-dependent second order reduced density matrix ($\rho^{(2)}$)~\cite{SakmannPRA2008, BeraSciRep2019} as a function of the spatial coordinates, which helps in obtaining insights into the evolving pair-correlations in real space. Here we focus on the evolution of the unnormalized pair density $\rho^{(2)}(i,j;t)$, which is related to the two-body correlation function $g^{(2)}(i,j;t)$ as
\begin{align}
  g^{(2)}(i,j;t)=\frac{\rho^{(2)}(i,j;t)}{\rho^{(1)}_{ii}(t)\rho^{(1)}_{jj}(t)}. 
\end{align}

The spatially resolved second order reduced density matrix
$\rho^{(2)}(i,j;t)$ is computed as 
 \begin{align}
  \rho^{(2)}(i,j;t)&=\rho^{(2)}_{ijji}(t)=\langle i,j|\hat{\rho}^{(2)}(t)| i,j\rangle\nonumber\\
  &= \sum_{a b c d}
\rho^{(2)}_{abcd}(t)\phi_{ai}^*(t)\phi_{bj}^*(t)\phi_{cj}(t)\phi_{di}(t),
\end{align}
 where $i,j$ are the discrete site indices and $a,b,c,d$ are discrete mode indices. Here, the system is initialized in the same way as in Fig.~\ref{mbfig3}.  The cross-peaks in the pair density map (the relevant observable for the analysis) are absent for a single-mode ($M = 1$) dynamics (Fig.~\ref{mbfig4}(a)-(d)) but are captured when going beyond the mean-field limit  ($M > 1$)  as shown in (Fig.~\ref{mbfig4}(e)-(h)) and (Fig.~\ref{mbfig4}(i)-(l)). With regard to the $M=2$ dynamics, the pair density signature is quite prominent  at 50 fs but thereafter decays, such that, at 480 fs, the earlier enhanced regions of pair densities are now barely visible, illustrating the transient nature of the pair correlations in an environment dominated by phonons. A similar behavior is reproduced when using  $M=3$ which suggests that $M=2$ produces reasonably converged results for the set of parameters chosen here.  Our numerical results thus illustrate the limitation of the single-mode (mean field) description, especially in reproducing spatial pair density signatures, even though the same mean-field description provides a semi-quantitatively accurate diffusion constant.

\begin{figure*}[!t]
\centering
\includegraphics[width=1.0\linewidth]{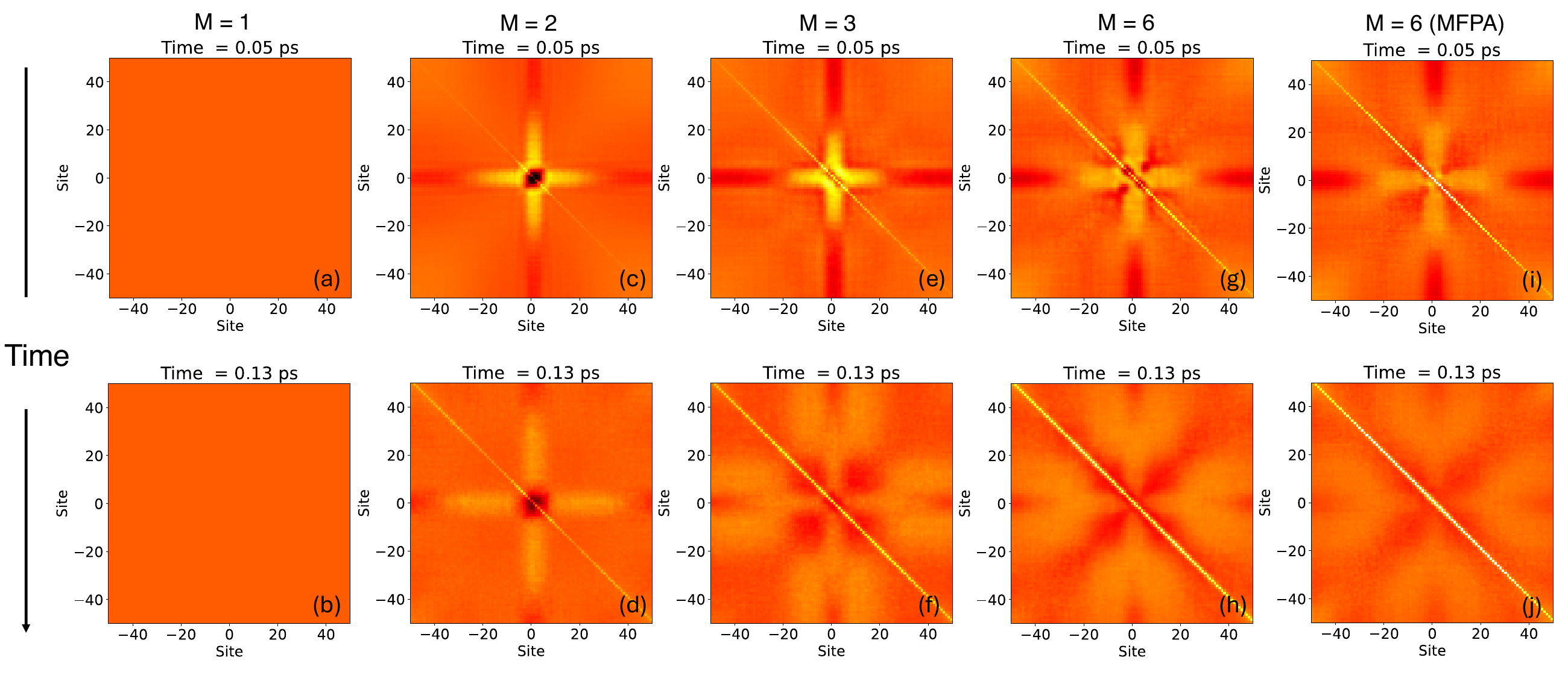}
\caption{\footnotesize 
Time-dependent spatial pair correlation function $g^{(2)}$, plotted at two representative times: $0.05$ ps (top row) and $~0.13$ ps (bottom row), well within the early-time window of the many-body exciton dynamics, where $N_\mathrm{ex}=10$ and $U =0.01$ a.u., for $M=1,~2,~3,~6$ and $M=6$ (MFPA). The rest of the parameters are identical to those in Fig.~\ref{mbfig3}.}
\label{mbfig6}
\end{figure*}
In Fig.~\ref{mbfig5}, we analyze the importance of accurately describing nuclear correlated nonadiabatic forces. Here we compare the relative performance of three possible approximations for propagating the correlated exciton-phonon dynamics. We compute and compare the MSDs  with $M=3$ and $U =0.01$ a.u. and implement  three specific approximations.  Fig.~\ref{mbfig5}(a) shows the {\it mean field path approximation} (MFPA), where the nuclear trajectories are precomputed using  $M=1$ dynamics. This is in the same spirit as in the classical path approximation, which is often implemented in simulating exciton or polaritonic dynamics, especially in the single excitation/low excitation limit~\cite{MandalJPCA2020, MattosJCTC2024, ChngNL2025, akimov2013pyxaid,wang2011mixed}.
Specifically, we use 
\begin{align}
    \{R_i(t)\},\{P_i(t)\} \equiv \{R_i(t)\}^{ref}_{M=1},\{P_i(t)\}^{ref}_{M=1}.
\end{align}
We find that this approximate propagation  does not reproduce  the MSD, which deviates from our benchmark calculation performed with $M=3$. This illustrates that the overall dynamics is sensitive to the nuclear back-reaction. 

In Fig.~\ref{mbfig5}(b) we explore a different approximation, namely a {\it mean field force approximation}, where the nuclei are propagated with the nonadiabatic forces computed  using the mean-field expression  
\begin{align}
  \dot{P}_i(t)   = -\omega_0^2R_i(t) - \gamma[\rho^{(1)}_{ii}(t)]_{M=1},
\end{align}
 while the excitonic subsystem is evolved in a fully correlated manner. The result of such an approach is presented in Fig.~\ref{mbfig5}(b), which shows that this approach also fails to reproduce accurate dynamics.    
 
Finally, in Fig.~\ref{mbfig5}(c) we explore a commonly used approximation, namely, the {\it classical path approximation}, where the  nuclear trajectories  are evolved in the ground state potential energy surface, such that:
 \begin{align}
\dot{P}_i(t) &= -\omega_0^2R_i(t).
 \end{align}

This means, while the nuclear back reaction is neglected,  the usual parametric dependence of the one-body Hamiltonian on the nuclear coordinates is still retained. We find that, just like its predecessors, it fails to reproduce correlated excitonic diffusion.  

Fig.~\ref{mbfig5}(d) records the average value of the two-body reduced density matrix $\rho^{(2)}(x_1,x_2;t)$ within a fixed square window ($W$) centered at $(0,-10)$ and having a half-width of $5$ units in the discrete two-body coordinate space, for the reference and the approximate approaches considered. Here, the designated location of the window has been marked using a black square in all the correlation maps shown subsequently (Fig.~\ref{mbfig5}(e)-(l)). Early-time  propagation shows an initial enhancement in the average spatial pair density, indicating the existence of transient spatial two-body correlations, as expected from the results discussed in Fig.~\ref{mbfig4}, which is absent in the $M=1$ propagation. The pair density maps for the MFPA propagation (Fig.~\ref{mbfig5}(i)-(l)) relative to the reference $M=3$ dynamics (Fig.~\ref{mbfig5}(e)-(h)) shows that it can capture the nuances of the pair density to reasonable accuracy. Out of all the three approximation approaches considered, we find that only the mean-field path approximation (MFPA) exhibits reasonably consistent two-body reduced density matrix, for the entire duration of time considered.  Thus, despite being insufficient in capturing the long time exciton transport dynamics,  the MFPA approach consistently computes the early time spatial pair densities utilizing precomputed trajectories from the reference $M=1$ simulation, making this a computationally efficient substitute suitable for investigating many-body spatial correlations at sub-picosecond timescales.

Finally, in Fig.~\ref{mbfig6} we present the direct spatial correlations $g^{(2)}$. We find that $M = 6$ produces converged $g^{(2)}$ which are presented in Fig.~\ref{mbfig6}(g)-(h). Fig.~\ref{mbfig6}(a)-(b), which presents the results when setting $M=1$, shows that a mean-field treatment fails to capture any spatial correlations, as expected. In comparison, $M=2$ or $3$, presented in Fig.~\ref{mbfig6}(c)-(f), produces qualitatively accurate spatial correlations. Finally, we find that MFPA, presented in Fig.~\ref{mbfig6}(i)-(j), accurately reproduces the reference converged correlations ($M=6$) which corroborates our findings in Fig.~\ref{mbfig5}. 

\section{Conclusion}
In this work, we have developed a multiconfigurational mixed quantum-classical approach for simulating  many-body quantum dynamics of interacting excitonic systems in the presence of phonon-induced dynamic disorder. The proposed method systematically incorporates multiple modes, allowing us to go  beyond a mean-field (or single-mode) description of the many-body wavefunction, while the nuclei propagate quasi-classically according to the MTE framework through exciton-phonon coupling.  This unified algorithm is also shown to capture the transient spatial pair density signatures of multiple interacting excitations in the presence of phonon-induced effects incorporated by static and dynamic disorder, and it exhibits progressive convergence toward the exact dynamics of the quantum (excitonic) subsystem as the number of modes is increased, which we have confirmed through benchmarking against exact quantum dynamics in reduced model systems. Our results also show that the emergence of many-body correlations (as indicated by the enhanced pair densities at early times) affect the transport dynamics in such excitonic systems,  modifying the diffusive behavior of excitonic transport. Additionally, we assess the feasibility of certain approximation schemes and their relative accuracies with respect to our reference multiconfigurational dynamics. Although these approximate approaches fail to reproduce the reference transport dynamics in a multi-excitonic system, we find the mean-field path approximation (MFPA) to be particularly useful in reproducing early-time spatial pair densities that are in close agreement with the reference multiconfigurational exciton dynamics.

Overall, our method overcomes the drawbacks of our prior many-body mixed quantum-classical approach which was limited to the mean-field limit, thereby enabling robust investigations of many-body correlations in interacting multi-excitonic systems. This methodology  is expected to serve as a useful platform for studying correlated many-body phenomena in more complex systems, including transport and relaxation dynamics in systems with strong light-matter coupling.
\section{Acknowledgments}
This work was supported by the Texas A\&M startup funds and  by the
U.S. National Science Foundation (NSF) under Grant
No. CHE-2611431. This work used TAMU ACES and LAUNCH clusters at the Texas A\&M University through allocation  PHY250275 and PHY230021 from the Advanced Cyberinfrastructure Coordination Ecosystem: Services \& Support (ACCESS) program, which is supported by National Science Foundation grants \#2138259, \#2138286, \#2138307, \#2137603, and \#2138296.  The authors appreciate discussions with Logan Blackham, Saeed Rahmanian Koshkaki, Sachith Wickramasinghe, Arshath Manjalingal, Amir Amini and Michael Fowler. 
\bibliography{bib.bib}
\end{document}